\documentclass[10pt, conference]{IEEEtran}
\usepackage{cite}
\usepackage{amsmath,amssymb,amsfonts}
\usepackage{graphicx,setspace}
\usepackage{textcomp}
\usepackage{algorithm,algpseudocode}
\newcommand\CONDITION[2]%
  {\begin{tabular}[t]{@{}l@{}l@{}}
     #1&#2
   \end{tabular}%
  }
\algdef{SE}[WHILE]{While}{EndWhile}[1]%
  {\algorithmicwhile\ \CONDITION{#1}{\ \algorithmicdo}}%
  {\algorithmicend\ \algorithmicwhile}
\algdef{SE}[FOR]{For}{EndFor}[1]%
  {\algorithmicfor\ \CONDITION{#1}{\ \algorithmicdo}}%
  {\algorithmicend\ \algorithmicfor}
\algdef{S}[FOR]{ForAll}[1]%
  {\algorithmicforall\ \CONDITION{#1}{\ \algorithmicdo}}
\algdef{SE}[REPEAT]{Repeat}{Until}{\algorithmicrepeat}[1]%
  {\algorithmicuntil\ \CONDITION{#1}{}}
\algdef{SE}[IF]{If}{EndIf}[1]%
  {\algorithmicif\ \CONDITION{#1}{\ \algorithmicthen}}%
  {\algorithmicend\ \algorithmicif}%
\algdef{C}[IF]{IF}{ElsIf}[1]%
  {\algorithmicelse\ \algorithmicif\ \CONDITION{#1}{\ \algorithmicthen}}
\usepackage{textcomp}
\usepackage{array}
\usepackage{tabularx}
\usepackage{booktabs}
\usepackage{mathtools}
\usepackage{comment}
\usepackage{amsthm}
\usepackage{url}
\usepackage{siunitx}
\usepackage{pgfplots}\pgfplotsset{compat=1.9}
\usepackage{pgfplots,capt-of}

\newcommand{\squeezeup}{\vspace{-2.5mm}}
\usepackage{scalefnt}
\usepackage{bigints}
\usetikzlibrary {arrows.meta}
\usepackage{epstopdf}

\usepackage{subfigure}
\usepackage[font=small,skip=0pt]{caption}
\newtheorem{theorem}{Theorem}

\theoremstyle{theorem}
\begin{document}
\title{Merits of Serving UAVs via Terrestrial Networks: A Vertical Antenna Radiation Study
\thanks{}}
\author{

\IEEEauthorblockN{  Nesrine~Cherif\IEEEauthorrefmark{1} and  Qurrat-Ul-Ain Nadeem\IEEEauthorrefmark{1}\IEEEauthorrefmark{2}\\
Email:	\!\{nc3654, qurrat.nadeem\}@nyu.edu\\
	\IEEEauthorblockA{\IEEEauthorrefmark{1} Engineering Division, New York University (NYU), Abu Dhabi, UAE\\
 \IEEEauthorrefmark{2}  NYU Tandon School of Engineering, New York, USA
		}
		}
}

\maketitle
\linespread{1.1}
\begin{abstract}
Unmanned Aerial Vehicles (UAVs) are increasingly used in a plethora of applications such as shipping, surveillance, and search-and-rescue. For UAVs to operate safely, reliable cellular connectivity is essential. Utilizing the terrestrial networks for aerial connectivity has been proposed, but the 3D radiation pattern of base station antennas significantly affects the performance of aerial links.. To address this, we evaluate the coverage probability of cellular-connected UAVs, considering vertical antenna gain, by leveraging tools from stochastic geometry. We also analyze how the UAV hovering height, tilt angle and 3D antenna beamwidth influence the reliability of the communication link. Our results show that a down-tiled antenna  does not only improve the connectivity of terrestrial users but also its cellular-connected UAVs counterpart. Moreover, the coverage probability of the UAV-UE becomes saturated
at large down-tilt angles at the TBSs due to the antenna sidelobe gain at the serving and interfering TBSs. We also found that the significant increase of the vertical antenna beamwidth improves the UAV user coverage probability especially at relatively low hovering altitudes thanks to the increase of the desired signal strength compared to the interference power.

\end{abstract}

\section{Introduction}
Unmanned aerial vehicle (UAV) based applications have proliferated recently. More specifically, UAVs are deployed in multiple scenarios from video surveillance to smart irrigation and agriculture, search-and-rescue mission and last mile cargo delivery \cite{cao2018airborne,Hayat}, to name a few. Thus, reliable cellular connectivity is of paramount importance for the successful implementation of these applications. This is driven not only by the need for high-quality service (QoS) but also by strict safety regulations to protect individuals and property beneath the UAVs. These stringent guidelines demand robust cellular link to ensure reliable operation, prevent accidents, and maintain control even in challenging conditions. To enable the fast mass deployment of cellular-connected UAV user equipment (UAV-UE), re-utilizing the terrestrial cellular network presents as the fastest and most  cost-effective solution for providing cellular coverage to these users at elevated heights. In fact, the merit of delivering cellular connectivity to aerial users, such as UAV-UEs, has been discussed in 3GPP technical report \cite{3gpp777}, which enumerated the main challenges around the coexistence of  aerial users with  terrestrial networks that include poor signal quality in elevated heights coupled with high interference from terrestrial base stations (TBSs) due to the strong line of sight (LoS) links.

Re-utilizing the current terrestrial networks that were primarily designed for terrestrial users  with an average height of 1.5 meters, for servicing aerial users at heights of tens of meters is a challenging idea \cite{mozaffari2018tutorial}. This is mainly due to the down-tilt of antennas at the TBSs that has been typically designed to uniquely serve users beneath the radio tower, i.e., terrestrial users \cite{chen32023}. As a consequence, the vertical radiation pattern of TBS antennas will have a substantial impact on the feasibility of using terrestrial networks to deliver reliable cellular connectivity to aerial users. Moreover, the fundamental difference in the propagation environment between the TBSs and aerial users and the TBSs and ground users lies in the effect of blockages and shadowing that reduces significantly at elevated altitudes \cite{cherifdownlink}. More specifically, multiple works have studied the system model with LoS and non-LoS (NLoS) propagation \cite{azari,cherifdownlink}, and showed the impact of the hovering altitude of the UAV on  its probability of coverage. These works highlight the great impact of interference in LoS aerial propagation. However, they do not accurately quantify  the impact of TBS vertical antenna radiation pattern on  elevated altitudes and rather account for the mainlobe and sidelobe gains only. As a result, the exact vertical antenna radiation pattern, parametrized by the down-tilt and half power beamwidth (HPBW), has been omitted from the coverage analysis of the UAV user equipment (UAV-UE) in these works. On the other hand, the authors in \cite{aquino2015} studied the optimal vertical tilt angle in a heterogeneous terrestrial network comprising of marco and femto cells. This work highlighted the significant influence of the vertical antenna down-tilt on the coverage of the terrestrial user as well as on the severity of  interference experienced by it. This motivates us to analyze and study quantitatively the joint effect of the TBS  down-tilt angle, the half power beamwidth, i.e., 3dB beamwidth, as well as the altitude of the UAV among other network parameters on the coverage performance of a typical UAV-UE.

The main purpose of this paper is to study the framework comprising of a cellular terrestrial network and UAV-UEs, and analyze the impact of different parameters affecting the integration of this new type of user at elevated heights. The parameters impacting the performance of the UAV-UE are expected to include the TBS antenna down-tilt and beamwidth as well as the hovering altitude of the UAV-UE. By leveraging tools from stochastic geometry, we quantify the coverage probability as a sum of different integrals, where the range of each integral is highly dependent on the difference of altitudes between the TBSs and the typical UAV-UE, and the vertical antenna radiation parameters, i.e., down-tilt and 3dB beamwidth. The results show that the values of down-tilt and beamwidth have a significant impact on the relative strength of the desired signal at the UAV-UE from the serving TBS, compared to the interference from all remaining TBSs. An optimal antenna tilt can be set and is found numerically to maximize the coverage probability for both the aerial and terrestrial users.

The remainder of the paper is organized as follows. In Section II, we detail the system model including the vertical antenna radiation pattern and the channel model. Section III details the coverage probability analysis. An extensive set of numerical results is provided in Section IV. Finally, we conclude this work in Section V.
\squeezeup
\section{System Model}
\label{sec:sysmodel}
We assume a terrestrial network $\Phi$ comprising of uniformly distributed TBSs, i.e., homogeneous Poisson point process (HPPP), with density $\lambda$ (number of TBSs/Km$^2$) \cite{aquino2015}. The height and transmit power of TBSs  are denoted by $h_{\rm BS}$ and $P_{\rm T}$, respectively, and assumed to be the same for all TBSs. In this work, the typical aerial user, i.e., UAV-UE, is hovering at an altitude $h_{\rm UAV}$ and assumed to be equipped with an omni-directional antenna. 
The antenna radiation pattern of the TBSs is assumed to be
horizontally omni-directional. Whereas, its vertical counterpart is assumed to be directional and parametrized by the antenna down-tilt and 3dB beamwidth.
Next, we give a brief overview of the vertical antenna pattern of the TBSs followed by the channel model between the TBSs and the typical UAV-UE.

\subsection{Vertical Antenna Pattern}
\begin{figure}[t!]
	\centering
	\includegraphics[scale=0.4]{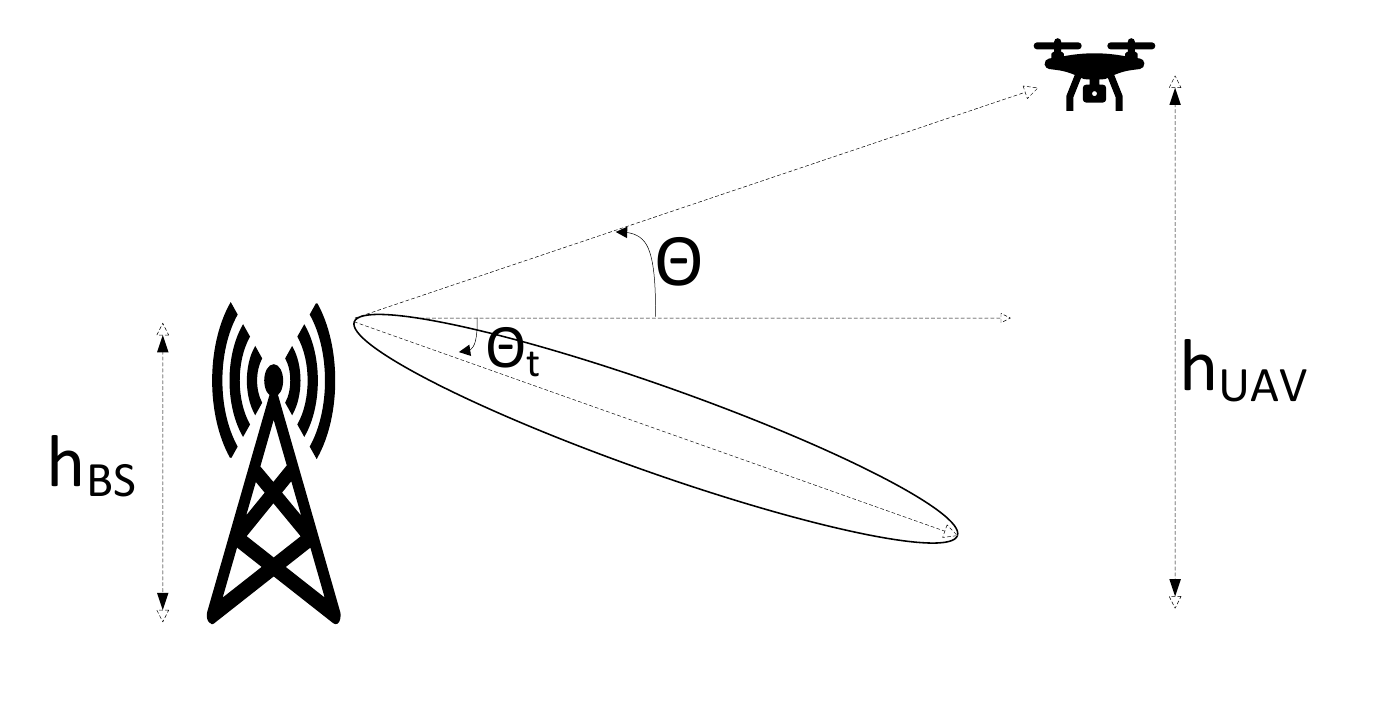}
	\caption{Illustration of 3D vertical angle between the UAV-UE and the TBS. }
	\label{fig:sys}
\end{figure}

The vertical antennas pattern in dB is expressed in \cite[Table A.2.1.1-2 ]{3gpp36814} as follows
\begin{equation}
    \label{eq:antennavertical}
    G_V(\theta)=-\text{min} \left[12\left(\frac{\theta-\theta_{\rm t}}{\theta_{\rm 3dB}}\right)^2\text{, }A_V\right],
\end{equation}
where $\theta$ is the elevation angle between the TBS and the UAV-UE as shown in Fig. \ref{fig:sys}, $\theta_{\rm t}$ and $\theta_{\rm 3dB}$ are the antenna down-tilt and 3dB beamwidths respectively, and $A_V$ is the minimum vertical gain that typically has a value of 20 dB. 

\begin{figure}
	\centering
	\includegraphics[width=0.9\linewidth]{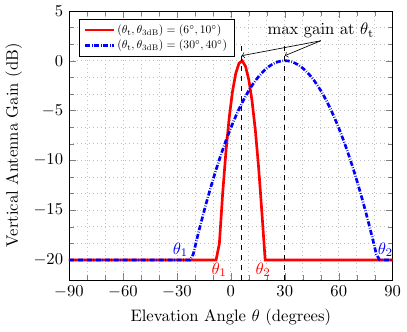}
\captionof{figure}{Vertical antenna radiation pattern of TBS.}
\label{fig:elevation}
\end{figure}

The elevation angle $\theta$ for the UAV-UE can be rewritten as
\begin{equation}
    \label{eq:elev}
    \theta = -\text{sin}^{-1}\left(\frac{h_{\rm d}}{r}\right),
\end{equation}
where $h_{d}=h_{\rm UAV}-h_{\rm BS}$ is the difference in altitudes between the TBS and the UAV-UE, and $r$ is the 3D distance between the TBS and the UAV-UE. In Fig. \ref{fig:elevation}, we plot the vertical antenna gain in dB with respect to the elevation angle in [-90$^\circ$, 90$^\circ$]. It is worth mentioning that for a terrestrial user, the elevation angle is always positive since the height of the terrestrial user is smaller than its TBS counterpart. Whereas for the UAV-UE, the elevation angle is always negative as the UAV-UE is expected to hover at altitudes higher than the TBSs height. Fig. \ref{fig:elevation} shows that the maximum vertical antenna gain can be attained when the UE is at an elevation angle equal to the tilt angle $\theta_{\rm t}$, and it starts decreasing at the same pace as the UE moves in both directions. This means that the strength of the vertical radiation pattern in the direction where the UAV-UE is located will depend on the down-tilt angle and the 3dB beamwidth set at the TBS. The vertical antenna gain attains the minimum value $A_V$ at two values of elevation angle, $\theta_1$ and $\theta_2$. Thus, we can rewrite the vertical antenna  pattern formula in \eqref{eq:antennavertical} as

\vspace{-0.5cm}

\begin{equation}
\label{eq:gain}
  G_{\theta_{\rm t},\theta_{\rm 3dB}}^{V}(\theta)=\left\{
\begin{array}{ll}
-A_V ,&\theta<\theta_1\\
-12\left(\frac{\theta-\theta_{\rm t}}{\theta_{\rm 3dB}}\right)^2, &\theta_1\leqslant\theta<\theta_2\\
-A_V ,&\theta\geqslant\theta_2,
\end{array}
\right.
\end{equation}
where $\theta_1=-\sqrt{\frac{A_V}{12}}\theta_{\rm 3dB}+\theta_{\rm t}$ and $\theta_2=\sqrt{\frac{A_V}{12}} \theta_{\rm 3dB}+\theta_{\rm t}$. 

Note that the elevation angle between the UAV-UE and the TBS is always a negative angle in the range $[-90^{\circ}, 0^{\circ}]$, where an elevation angle of $0^{\circ}$ is attained when the 2D distance between the UAV-UE and the TBS goes to infinity and an elevation angle of $-90^{\circ}$ means that the UAV-UE is placed at the top of the TBS with a distance of $h_{\rm d}$. As a result, the vertical antenna gain for the UAV-UE can be defined based on the value of $\theta_1$. By plugging  (\ref{eq:elev}) into  (\ref{eq:gain}), we can express the vertical antenna pattern, in linear scale, as a function of the 3D distance $r$ between the TBS and the UAV-UE as

\begin{equation}
\label{eq:gain_r}
 G_{\theta_{\rm t},\theta_{\rm 3dB}}^{V}(r)=\left\{
\begin{array}{ll}
     \textbf{if } \underline{\theta_1 \geqslant 0^{\circ}:} &\\
     \quad\quad\quad\quad\quad A_V^L  \\
      \textbf{if } \underline{\theta_1 \leqslant 0^{\circ}:} & \\\\
     \begin{array}{ll}
A_V^L ,&r<r_1\\
10^{-1.2\left(\frac{-\sin^{-1}\left(\frac{h_{\rm d}}{r}\right)-\theta_{\rm t}}{\theta_{\rm 3dB}}\right)^2}, &r_1\leqslant r,
\end{array} 
\end{array}
\right.
\end{equation}
% By plugging eq. (\ref{eq:elev}) into eq. (\ref{eq:gain}), we can express the vertical antenna gain, in linear scale, as a function of the 3D distance between the TBS and the UAV-UE
% \begin{equation}
% \label{eq:gain_r}
%  G_{\theta_{\rm t},\theta_{\rm 3dB}}^{V}(r)=\left\{
% \begin{array}{ll}
% A_V^L ,&r<r_1\\
% 10^{-1.2\left(\frac{-\sin^{-1}\left(\frac{h_{\rm d}}{r}\right)-\theta_{\rm t}}{\theta_{\rm 3dB}}\right)^2}, &r_1\leqslant r <r_2\\
% A_V^L ,&r\geqslant r_2,
% \end{array}
% \right.
% \end{equation}
where $r_1 = \frac{h_{\rm d}}{\sin(-\theta_1)}$, and $A_V^L$ is the minimum gain $-A_V$ in linear scale. Please note that $r_1$ is a valid distance, i.e., it has a  positive value, because $\theta_1$ is negative, i.e., $\sqrt{\frac{A_V}{12}}\theta_{\rm 3dB}>\theta_{\rm t}$ \cite{aquino2015}.

\subsection{Channel Model}
For simplicity,  we assume that the channel between the TBS and the UAV-UE  is  LoS dominated. In fact, it has been proven in \cite{cherifdownlink} that  the LoS assumption is a tight lower bound to the downlink coverage performance of LoS/NLoS channel model. Moreover, the assumption becomes more justifiable when the UAV-UE hovers at significant altitudes \cite{cherifdownlink}. We assume that the channel undergoes Nakagami-m fading. Thus, the received power gain from the $i$-th TBS at the typical UAV-UE is Gamma-distributed, i.e., $g_i\sim\text{Gamma}\left(m,\frac{1}{m}\right)$, where $m$ is the fading parameter, with probability density function expressed as
\vspace{-7pt}
\begin{equation}
    \label{eq:pdf}
    f_{g_i}(x)=\frac{m^m x^{m x}}{\Gamma(m)} e^{-m x},
\end{equation}
where $\Gamma(.)$ is the Gamma function. Hence, the received power from the $i$-th TBS at the typical UAV-UE is expressed as 
\begin{equation}
    \label{eq:pr}
    P_i^r=P_{\rm T}  G_{\theta_{\rm t},\theta_{\rm 3dB}}^{V}(r_i) g_i r_i^{-\alpha},
\end{equation}
with $\alpha$ is the path loss exponent and $r_i$ denotes the 3D euclidean distance between the $i$-th TBS and the typical UAV-UE.

The typical UAV-UE is assumed to associate with the closest TBS, $x_0^*$ \cite{cherifdownlink}, as follows
\vspace{-7pt}
\begin{equation}
    \label{eq:association}
    x_0^*=\min_{x_i \in \Phi}(r_i).
\end{equation}
Since thermal noise  power is negligible  compared to the interference power\cite{azari}, we assume that the UAV-UE operates in an interference-limited environment. Thus, the downlink signal to interference ratio $\gamma$ at the UAV-UE is given by
\begin{equation}
\label{eq:SIR}
    \gamma = \frac{P_{\rm T}  G_{\theta_{\rm t},\theta_{\rm 3dB}}^{V}\!\!(r_0) g_0 r_0^{-\alpha}}{I},
\end{equation}
where $r_0$ is the distance between the UAV-UE and its closest serving TBS and $I$ represents the interference received at the typical UAV-UE and is expressed by
\begin{equation}
    \label{eq:interference}
    I= \sum_{x_i\in \Phi \setminus x_0^*}P_{\rm T} G_{\theta_{\rm t},\theta_{\rm 3dB}}^{V}\!\!(r_i) g_i r_i^{-\alpha}
\end{equation}
\section{Coverage Probability}
The coverage probability $P^{\rm cov}$ is defined as the probability that the received SIR $\gamma$ at the UAV-UE is above a certain predetermined threshold $\beta_{\rm th}$, that describes the cellular requirement at the UAV-UE. Using eq. (\ref{eq:SIR}), we can express $P^{\rm cov}$ as
    \begin{eqnarray}
    \label{eq:pc}
    P^{\rm cov} &=& \mathbb{P}(\gamma\geqslant \beta_{\rm th})\nonumber\\
    &=&\sum_{k=0}^{m-1}\!\!\!\frac{(-1)^k}{k!}\left(m  \beta_{\rm th} \right)^k\nonumber\\
    &&\times \mathbb{E}_{r}\!\!\left[\frac{r^{k\alpha}}{G_{\theta_{\rm t},\theta_{\rm 3dB}}^{V}(r)^k} 
\left[\frac{\partial^k \mathcal{L}_I(s)}{\partial s^k}\right]_{s=\frac{m \beta_{\rm th}r^\alpha}{P_{\rm T} G_{\theta_{\rm t},\theta_{\rm 3dB}}^{V}(r)}} \right],\nonumber\\
\end{eqnarray}
where $\mathcal{L}_I(s) = \mathbb{E}_I\left(-sI\right)$ is the Laplace transform of the TBS interference.

\textit{Proof:} The proof is postponed to Appendix \ref{ap:cov}.

The Laplace transform $\mathcal{L}_I(s)$ of the aggregated  interference power received by the UAV-UE from the TBSs is given by 
\begin{equation}
\label{eq:Li}
\mathcal{L}_{I}(s)\!\!=\!\!
	\left\{
\begin{array}{ll}
     \textbf{if } \underline{\theta_1 \geqslant 0^{\circ}:} &\\
     \quad \exp\Bigg(\!\!-\!2\pi \lambda\!\Bigg[\displaystyle\int_{r}^{\infty}\!\!\!\left(\!\!1\!\!-\!\!\left(\!\frac{\!\!\!m}{m\!+\!s P_{\rm T} A_V^L t^{-\alpha}}\right)^{\!\!\!m}\right)\! t dt \Bigg]\Bigg) \\
      \textbf{if } \underline{\theta_1 \leqslant 0^{\circ}:} & \\
  \!\!\!\!\exp\Bigg(\!\!-\!\!2\pi \lambda\!\Bigg[\!\!\displaystyle\int_{r}^{r_1}\!\!\!\left(\!\!1\!\!-\!\!\left(\!\frac{\!\!\!m}{m\!+\!s P_{\rm T} A_V^L t^{-\alpha}}\right)^{\!\!\!m}\right)\! t dt\\
\!\!\!\!\!+\!\!\!\displaystyle\int_{r_1}^{\infty}\left(\!\!1\!\!-\!\!\left(\!\frac{\!\!\!m}{m\!+\!s P_{\rm T} 10^{\!\!\!\!-1.2\left(\frac{-\sin^{-1}\left(\frac{h_{\rm d}}{t}\right)-\theta_{\rm t}}{\theta_{\rm 3dB}}\right)^2} t^{-\alpha}}\!\right)^{\!\!\!m}\right)\!\!\!\!\!\! \nonumber\\
\times t dt\Bigg]\Bigg).
\end{array}
\right.
\end{equation}

\textit{Proof:} The proof is postponed to Appendix \ref{ap:li}.

Based on the results above, we provide an analytical expression for the coverage probability in terms of the network and TBS vertical antenna radiation pattern parameters in the following theorem.

\begin{theorem}
The coverage probability for a typical UAV-UE \label{eq:pcov_d} is given by eq. (\ref{eq:Pcov}) provided at the top of next page, where $f_{r_0}(r)$ is the probability density function (PDF) of the distance between the nearest TBS and the typical UAV-UE which is given by \cite[eq. (8)]{chetlur2019coverage}:
\begin{equation}
    \label{eq:pdf_r}
    f_{r_0}(r)= 2 \pi \lambda r \exp(-\lambda \pi (r^2-h_{\rm d}^2)), \quad r \geqslant h_{\rm d}.
\end{equation}

\begin{figure*}
    \begin{equation}
\label{eq:Pcov}
 P^{\rm cov}=\left\{
\begin{array}{ll}
     \textbf{if } \underline{\theta_1 \geqslant 0^{\circ}:} &\\
     \quad\quad\quad\quad\quad \displaystyle\sum_{k=0}^{m-1}\!\!\!\frac{(-1)^k}{k!}\left(m  \beta_{\rm th} \right)^k \displaystyle\int_{h_{\rm d}}^{\infty}\frac{r^{k\alpha}}{{A_V^L}^k} 
\left[\frac{\partial^k \mathcal{L}_I(s)}{\partial s^k}\right]_{s=\frac{m \beta_{\rm th}r^\alpha}{P_{\rm T} A_V^L}}f_{r_0}(r) dr  \\
      \textbf{if } \underline{\theta_1 \leqslant 0^{\circ}:} & \\
   \quad\quad\quad\quad\quad \displaystyle \sum_{k=0}^{m-1}\!\!\!\frac{(-1)^k}{k!}\left(m  \beta_{\rm th} \right)^k \displaystyle\int_{h_{\rm d}}^{r_1}\frac{r^{k\alpha}}{{A_V^L}^k} 
\left[\frac{\partial^k \mathcal{L}_I(s)}{\partial s^k}\right]_{s=\frac{m \beta_{\rm th}r^\alpha}{P_{\rm T} A_V^L}}f_{r_0}(r) dr + \sum_{k=0}^{m-1}\!\!\!\frac{(-1)^k}{k!}\left(m  \beta_{\rm th} \right)^k \\
\quad\quad\quad\quad\quad \times\displaystyle\int_{r_1}^{\infty}\frac{r^{k\alpha}}{{10^{\!\!\!\!-1.2 k \left(\frac{-\sin^{-1}\left(\frac{h_{\rm d}}{r}\right)-\theta_{\rm t}}{\theta_{\rm 3dB}}\right)^2}}} 
\left[\frac{\partial^k \mathcal{L}_I(s)}{\partial s^k}\right]_{s=\frac{m \beta_{\rm th}r^\alpha}{P_{\rm T} 10^{\!\!\!\!-1.2\left(\frac{-\sin^{-1}\left(\frac{h_{\rm d}}{r}\right)-\theta_{\rm t}}{\theta_{\rm 3dB}}\right)^2}}}f_{r_0}(r) dr \\
\end{array}
\right.
\end{equation}
\hrule
\end{figure*}

\end{theorem}
\textit{Proof:} Plugging eq. (\ref{eq:gain_r}) in eq. (\ref{eq:pc}) and averaging over the distance in eq. (\ref{eq:pdf_r}) yields the coverage probability formula in eq. (\ref{eq:Pcov}).

The closed-form expression of the UAV-UE downlink coverage probability encompasses the TBS vertical antenna radiation parameters including down-tilt and 3dB beamwidth angles, height of the TBS and UAV-UE, large scale fading parameters, and the network density. Therefore it only depends on statistical values. As a result, it can be used for performance analysis and optimization of different parameters without relying on time consuming Monte-Carlo simulations. Moreover any optimization performed based on this expression does not need to be re-performed every time the channel realization changes, but only when the channel statistics change which happens on a much slower time scale. In the next section, we use this expression to numerically find the optimal down-tilt at TBSs that maximizes the coverage probability for the aerial and terrestrial users.
\vspace{-7pt}
\section{Simulation Results}
The network parameters used in the simulation are summarized in Table \ref{tab:param} and are mainly typical values drawn from \cite{3gpp36814}. All simulation results are assumed to be using these parameters unless stated otherwise. In the considered simulation setup, the UAV-UE's elevation angle with respect to the TBSs in the network typically lies in the range $[-22^{\circ}, -1^{\circ}]$, implying that while the elevation angle with respect to serving TBS might be closer to the lower end of this range, the elevation angle with respect to interfering TBSs can be closer to the higher end of this range, making the UAV-UE fall in the beamwidth of the interfering TBSs even if these TBSs adopt a down-tilt. 

\begin{figure*}[t!]
							\begin{minipage}[t][6cm][t]{0.31\textwidth}
\centering
\includegraphics[width=\textwidth]{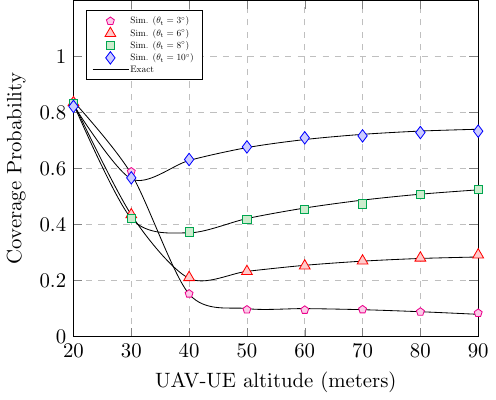}
\captionof{figure}{Coverage probability versus the UAV-UE altitude.}
\label{fig:cov_ht}
\end{minipage}
	\hspace{1em}
\begin{minipage}[t][6cm][t]{0.31\textwidth}
\centering
\includegraphics[width=\textwidth]{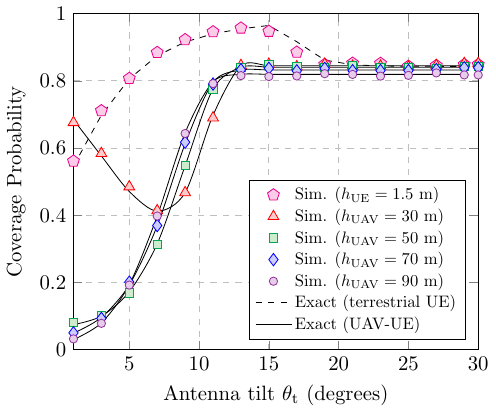}
\captionsetup{font=footnotesize}
\captionof{figure}{Impact of TBSs antenna down-tilt on  coverage probability.}
\label{fig:cov_tilt}
							\end{minipage}
       	           \hspace{1em}
							\begin{minipage}[t][6cm][t]{0.32\textwidth}
\centering
\includegraphics[width=\textwidth]{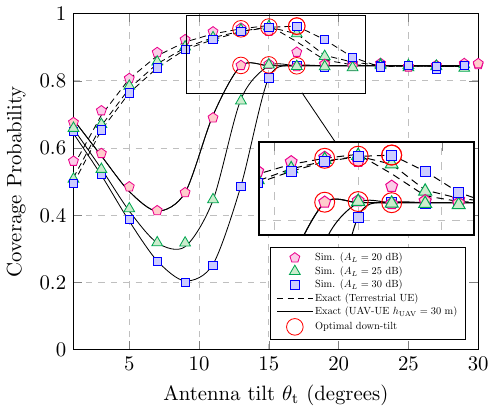}
\captionof{figure}{Impact of TBSs antenna down-tilt and sidelobe gain $A_V$ on  coverage probability.}
\label{fig:cov_tilt_Al}
\end{minipage}
\end{figure*}

In Fig. \ref{fig:cov_ht}, we plot the UAV-UE's coverage probability with respect to its altitude for different vertical antenna tilt angles at the TBSs. The closed-form expression of the coverage probability  in eq. (\ref{eq:Pcov}), marked as `Exact', matches very accurately the Monte Carlo based simulation results. For relatively small down-tilt angles, we observe that as the UAV-UE altitude increases, the coverage probability decreases until it reaches a minimum, and then it increases slightly. The decrease is due to the decrease in the desired signal strength with increasing altitude, while the eventual increase in coverage probability happens because as the altitude increases further, the interference experienced by the UAV-UE from non-serving TBSs decreases due to a larger path loss and sidelobe gain, and this effect dominates over the effect of decrease in desired signal strength causing the overall coverage probability to improve.  Moreover, counter-intuitively, we see that a larger down-tilt angle at the antennas of all TBSs yields a better coverage probability for the UAV-UE at elevated heights. This is because while a smaller down-tilt at the serving TBS increases the desired signal strength at the UAV-UE, at the same time a smaller down-tilt angle at all interfering TBSs  also results in higher unobstructed interference at the UAV-UE which deteriorates the SIR.
\begin{table}[!t]
	\caption{Network parameters}
	\label{tab:param}
	\centering
	\begin{tabular}{|c |c ||c| c|}
		\hline     Parameter &   Value   &   Parameter &   Value  \\  \hline
		$P_{\rm T}$   & 43 dBm     & $(m,\text{ }\alpha)$ & (2,\text{ }2.5)         \\ \hline
			$\beta_{\rm th}$   & -10 dB     & $h_{\rm BS}$ & 19 meters         \\ \hline
		$(\theta_{\rm t},\theta_{\rm 3dB})$   & $(6^{\circ},10^{\circ})$     & $\lambda $& 10 TBS/km$^2$ \ \\ \hline
 \hline
		
	\end{tabular}
 % \squeezeup
 \end{table}
% \vspace{15pt}

%%remaining three figures

\begin{figure*}[t!]
\begin{minipage}[t][6cm][t]{0.31\textwidth}
\centering
\includegraphics[width=\textwidth]{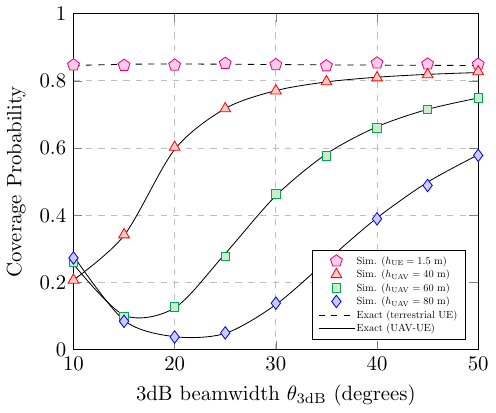}
\captionsetup{font=footnotesize}
\captionof{figure}{Impact of the  3dB beamwidth $\theta_{\rm 3dB}$ of TBS antennas on the coverage probability.}
\label{fig:cov_beam}
\end{minipage}
	\hspace{1em}
\begin{minipage}[t][6cm][t]{0.31\textwidth}
\centering
\includegraphics[width=\textwidth]{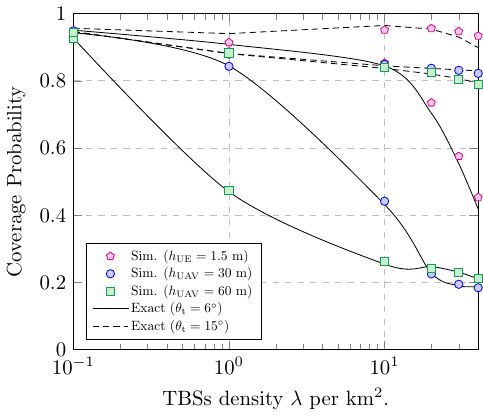}
\captionsetup{font=footnotesize}
\captionof{figure}{Coverage probability versus TBS density.}
\label{fig:cov_density}
							\end{minipage}
       	           \hspace{1em}
							\begin{minipage}[t][6cm][t]{0.31\textwidth}
\centering
\includegraphics[width=\textwidth]{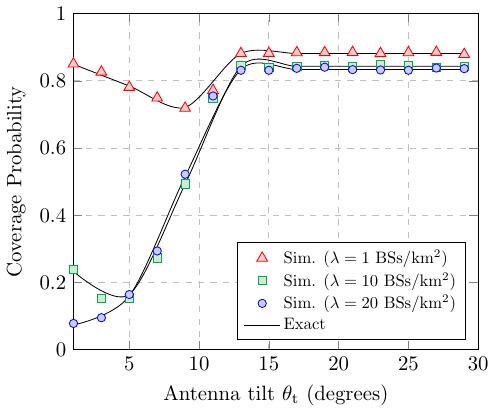}
\captionof{figure}{Impact of down-tilt of TBSs antennas on the coverage probability for $h_{\rm UAV}=40$ m and different TBS densities.}
\label{fig:cov_tilt_density}
\end{minipage}
\end{figure*}
We study the vertical antenna down-tilt angle impact on the coverage performance of a terrestrial user  and UAV-UE in Fig. \ref{fig:cov_tilt}. The dashed line plot the coverage probability for the terrestrial UE at a height of $1.5$ m, while the solid lines represent the performance of the UAV-UE at different altitudes. The terrestrial UE is assumed to be served in a similar manner with one serving TBS and other TBSs causing interference. The coverage probability of the UAV-UE  becomes saturated and oblivious to the UAV-UE altitude at larger down-tilt angles at the TBSs. This is because at larger down-tilt the  radiation pattern of the serving and interfering TBSs at the elevation angle of the UAV-UE is equal to the sidelobe gain $-A_V$, and thus does not depends on the down-tilt angle. Besides, an optimal down-tilt angle, that should be adopted by all TBSs, can be found that maximizes both the terrestrial UE's coverage and the UAV-UE's coverage at any UAV-UE height. This optimal down-tilt angle for the network under consideration is $\theta_{\rm t}^*=13^{\circ}$. Interestingly, the coverage probability of the UAV UE is reasonable even though the TBSs are down-tilting because of the reduced interference from interfering TBSs under down-tilting. It is important to point that the insights depend on the sidelobe level $A_V$ considered in the 3GPP model. Next, we fine tune the sidelobe gain $-A_V$ to investigate its impact on the overall performance of the UAV-UE and the optimal down-tilt angle.

We highlight the impact of the 3GPP vertical antenna radiation model on UAV-UE performance in Fig. \ref{fig:cov_tilt_Al}. Specifically, as the sidelobe gain  $-A_V$ decreases, the optimal down-tilt angle shifts slightly to the right for both terrestrial UE and UAV-UE. Additionally, with a smaller sidelobe gain, the coverage probability generally deteriorates until the optimal down-tilt angle is reached. This effect occurs because a smaller angle $\theta_1$ expands the region where the UAV-UE is within the mainlobe (see Fig. \ref{fig:elevation}), resulting in higher interference power which leads to a reduced UAV-UE coverage performance. Thus, we can conclude that the UAV-UE performance is highly sensitive to the 3GPP vertical antenna gain model in \cite{3gpp36814}.

In Fig. \ref{fig:cov_beam}, we investigate the coverage probability versus the  vertical antenna 3dB beamwidth for both terrestrial UE and UAV-UE. The probability of coverage for the terrestrial UE is steady with respect to the TBS antenna beamwidth. This is because the terrestrial UE is oblivious to the beamwidth in the vertical plane. Whereas for the UAV-UE, at $\theta_{\rm 3dB}=10^{\circ}$, we observe that the lower the UAV-UE height, the lower the probability of coverage which corroborates the results shown in Fig. \ref{fig:cov_ht}. Whereas at larger $\theta_{\rm 3dB}$, we find better performance for lower UAV-UE altitude. This outcome is mainly due to the fact that as the 3dB beamwidth increases, the UAV-UE becomes more likely to fall within the mainlobe of the radiation pattern of the serving BS, while due to larger distance from interfering TBSs, it does not experience a large increase in the received radiation and hence interference from them even with an increase in their beamwidth. This effect becomes more pronounced at lower altitudes where the UAV-UE can quickly get into the mainlobe of the serving TBS vertical radiation pattern.

The TBS density impact on the probability of coverage for both terrestrial and UAV UE is investigated in Fig. \ref{fig:cov_density}. As expected, overall, a higher TBS density results in higher interference power and thus a lower probability of coverage. Moreover, the decrease  is more explicit with smaller antenna down-tilt which corroborates the trend seen in Fig. \ref{fig:cov_ht}. As the TBS vertical antenna down-tilt angle increases, the impact of interference power on the performance decreases, particularly in denser environments., i.e., high $\lambda$. This can be explained by the fact that denser environment increases the likelihood that the serving TBS for the UAV-UE is at closer distance and thus improves the desired signal power. On the other hand, the interference also increases but at smaller rate with larger down-tilt angle. The terrestrial user coverage performance exhibits similar trend as its UAV-UE counterpart with respect to the TBS density.

 In Fig. \ref{fig:cov_tilt_density}, we investigate the impact of the terrestrial network density on the optimal down-tilt angle that maximizes the UAV-UE coverage probability. For relatively sparse terrestrial network, i.e., smaller $\lambda$, there exists a down-tilt angle that minimizes the UAV-UE coverage probability. This angle increases as the density of the terrestrial network decreases. This can be explained by noting that smaller down-tilt can provide better performance in low density networks since interference is small and improving the desired signal at the UAV-UE using a smaller down-tilt has a profound impact. However as the network density increases and interference becomes dominant, it is always beneficial to increase the down-tilt angle up to the optimal value to decrease interference. Moreover, the probability of coverage saturates at the optimal down-tilt angle corroborating the trend seen in Fig \ref{fig:cov_tilt}. Interestingly, the optimal down-tilt angle that maximizes the UAV-UE coverage remains the same (around $13^{\circ}$) under different terrestrial network density or UAV-UE hovering altitude.

\section{Conclusion}
We quantified accurately the probability of coverage of a typical UAV-UE served by a network of TBSs accounting for the vertical antenna radiation pattern using stochastic geometry. Numerical results showed that a smaller antenna down-tilt angle decreases the coverage probability at the UAV-UE considerably due to the significant increase of the interference power compared to the desired signal. Moreover, an optimal antenna down-tilt is found that simultaneously maximizes the coverage probability of the terrestrial user and the UAV-UE. Simulations showed that the coverage probability saturates at the optimal down-tilt angle thanks to the sidelobe antenna gain across both the serving and interfering TBSs. Overall, a notable increase in TBS antennas' 3dB beamwidth enhances the  coverage probability for the UAV-UE by improving the quality of the received desired signal more significantly than it increases the interference power.
\appendices

\vspace{-0.25cm}
\renewcommand{\thesection}{\Alph{section}}
 \section{}
 The derivation of $P^{\rm cov}$ in \eqref{eq:pc} is provided below.
\label{ap:cov}
\begin{eqnarray}
    \label{eq:pc_app}
    P^{\rm cov} &=& \mathbb{P}(\gamma\geqslant \beta_{\rm th})\nonumber\\
   &\overset{(a)}{=}&\mathbb{E}_{g_0,r_0,I}\left[\mathbb{P}\left(\frac{P_{\rm T}  G_{\theta_{\rm t},\theta_{\rm 3dB}}^{V}\!\!(r_0) g_0 r_0^{-\alpha}}{I}\geqslant \beta_{\rm th}\right)\right]\nonumber\\
   &\overset{(b)}{=}&\!\!\!\mathbb{E}_{r_0,I}\left[\mathbb{E}_{g_0}\left[\mathbb{P}\left(g_0\geqslant \frac{ r_0^{\alpha}\beta_{\rm th} I}{P_{\rm T}  G_{\theta_{\rm t},\theta_{\rm 3dB}}^{V}\!\!(r_0)}\right)\right]\right]\nonumber\\
    &\overset{(c)}{=}&\!\!\!\mathbb{E}_{r_0}\left[\mathbb{E}_{I}\left[\frac{\Gamma \left(m,\frac{m \beta_{\rm th} I r_0^{\alpha}}{P_{\rm T} G_{\theta_{\rm t},\theta_{\rm 3dB}}^{V}\!\!(r_0)}\right)}{\Gamma(m)}\right]\right],
\end{eqnarray}
where (a) follows by averaging the coverage probability over the random variables, i.e., the distance $r_0$, the channel gain $g_0$, and the interference $I$. Moreover, (b) follows from the independence between the random variables. The complementary cumulative distribution function of the Gamma distributed $g_0$ is applied in step (c).
Finally, by assuming an integer fading parameter $m$ and using the series expansion of the upper incomplete Gamma function, the formula in eq. (\ref{eq:pc}) follows after applying the definition of the Laplace transform of the interference $I$.

 \section{}
\label{ap:li}

The Laplace transform of the interference power received from TBSs at the UAV-UE is given by
\begin{eqnarray}
\label{eq:LIproof}
\!\!\!\!\!\!\!\!\!\!\!\!&&\!\!\!\!\!\!\!\!\!\!\!\!\mathcal{L}_{I}(s )=\mathbb{E}_{I}\left[\exp(-s  I) \right]\nonumber\\
\!\!\!\!\!\!\!\!\!\!\!\!&\overset{(a)}{=}&\!\!\!\!\!\!\mathbb{E}_{r_i}\left[\prod_{x_i\in\Phi \setminus x_0^*}^{}\mathbb{E}_{g_i}\left[\exp\left(-s P_{\rm T} g_i G_{\theta_{\rm t},\theta_{\rm 3dB}}^{V}(r_i) r_i^{-\alpha}\right)\right]\right]\nonumber\\
\!\!\!\!\!\!\!\!\!\!\!\!&\overset{(b)}{=}&\!\!\!\!\!\!\mathbb{E}_{r_i}\!\!\left[\prod_{x_i\in\Phi \setminus x_0^*}^{}\left(\frac{m}{m+s P_{\rm T} G_{\theta_{\rm t},\theta_{\rm 3dB}}^{V}(r_i)r_i^{-\alpha}}\right)^{m}\!\right]\\
\!\!\!\!\!\!\!\!\!\!\!\!&\overset{(c)}{=}&\!\!\!\!\!\!\exp\Bigg(\!\!-\!2\pi \lambda \int_{r}^{\infty}\!\!\!\left(\!\!1\!\!-\!\!\left(\!\frac{\!\!\!m}{m\!+\!s P_{\rm T} G_{\theta_{\rm t},\theta_{\rm 3dB}}^{V}(t) t^{-\alpha}}\right)^{\!\!\!m}\right)\! t dt\Bigg)\nonumber
\end{eqnarray}
where (a) follows from  (\ref{eq:interference}) and the independence of channel gain of the interfering TBSs and  their separating distances from the typical UAV-UE that follows a PPP. The moment generating function of the Gamma distribution was used in (b), and (c) is obtained  using the probability generating functional (PGFL). Plugging eq. (\ref{eq:gain_r}) into eq. (\ref{eq:LIproof}) yields eq. (\ref{eq:Li}).
\bibliographystyle{IEEEtran}  
\bibliography{references,Bibliography}

% This is due to the
% LOS propagation conditions, as at high altitudes, UAV
% capture more cells without obstructions, resulting in a stronger
% signal but with more interference. Interference mitigation is
% required to address this issue. The finding also revealed the
% practicality of using a dual network to improve the RSRQ of
% the UAV.

\end{document}